# Modulations in martensitic Heusler alloys originate from nanotwin ordering

M. E. Gruner[1,2], R. Niemann[2], P. Entel[1], R. Pentcheva[1], U. K. Rössler[2], K. Nielsch[2], S. Fähler[2] *

[1] *Theoretical Physics and Center for Nanointegration, CENIDE, University of Duisburg-Essen, D-47048 Duisburg, Germany*

[2] *IFW Dresden, Helmholtzstrasse 20, D-01069 Dresden, Germany*

**Abstract**

Heusler alloys exhibiting magnetic and martensitic transitions enable applications like magnetocaloric refrigeration and actuation based on the magnetic shape memory effect. Their outstanding functional properties depend on low hysteresis losses and low actuation fields. These are only achieved if the atomic positions deviate from a tetragonal lattice by periodic displacements. The origin of the so-called modulated structures is the subject of much controversy: They are either explained by phonon softening or adaptive nanotwinning. Here we used large-scale density functional theory calculations on the $Ni_2MnGa$ prototype system to demonstrate interaction energy between twin boundaries. Minimizing the interaction energy resulted in the experimentally observed ordered modulations at the atomic scale, it explained that *a*/*b* twin boundaries are stacking faults at the mesoscale, and contributed substantially to the macroscopic hysteresis losses. Furthermore, we found that phonon softening paves the transformation path towards the nanotwinned martensite state. This unified both opposing concepts to explain modulated martensite.

* corresponding author's email: S.Faehler@ifw-dresden.de

Ni-Mn-based Heusler alloys exhibit diffusionless phase transformations from a cubic high temperature austenite to a low temperature martensitic phase, which provide the opportunity for more energy efficient magnetocaloric refrigeration[1] as well as large stroke actuation by magnetically induced reorientation of twin boundaries.[2] They manifest outstanding functional magnetic properties, such as low transformation hysteresis losses[3] and very low actuation fields[4] only when the martensitic structure is modulated. In modulated structures the atomic positions deviate in a quite periodic way from a simple tetragonal structure. The origin of the modulation is not well-understood, with two competing explanations put forward, so far.[5-8]

The first explanation considers the electronic instability of the austenite phase. Fermi surface nesting within the cubic unit cell results in a soft phonon. This means that a lattice vibrational mode with a specific wave vector can be excited at a very low energy, which allows for an easy shearing of lattice planes.[9-11] In this picture, the modulations are interpreted as large shear movements that are frozen in the equilibrium martensitic phase.

The second explanation considers the phase boundary between cubic austenite and tetragonal martensite, which is required to form during a first order transition. To minimise the elastic strain energy at the phase boundary, twinning is necessary, i.e., the orientation of the tetragonal unit cells must alternate along the phase boundary. The different orientations of martensitic unit cells are connected by twin boundaries. If the elastic energy dominates over the twin boundary energy, it is beneficial to reduce the spacing of the twin boundaries to nanoscale. In the adaptive concept of Khachaturian et al.,[12] modulated martensite represents the shortest twinning periodicity, given by the lattice parameters. In Ni-Mn-Ga,[11,13,14] the modulated structure is considered to be metastable because microstructural defects like twin boundaries have an excess energy, which can drive a transition towards the tetragonal ground state.[15]

The adaptive concept is successful in describing the lattice constants of modulated martensite,[16] its magnetocrystalline anisotropy,[15] and the difference of twinning stress[13] and twin boundary energy[17] between modulated and tetragonal martensite. However, the concept has major shortcomings: it is not explaining why arrangements of nanotwin boundaries are periodic at all because irregular arrangements of nanotwin boundaries could also fulfil the elastic constraints at a phase boundary. Moreover, the preference of particular modulations, which are often stable with respect to temperature[18] and cycle number[19], cannot be derived from the concept. A particular shortcoming of the adaptive concept is that it cannot explain the absence of nanotwins that exhibit a width of only one lattice plane. But also the soft phonon concept has a major shortcoming. A soft phonon is a shear instability of {101} planes, but the energetically favoured transformation path follows part of the classical Bain transformation path[20], which occurs by elongation and compression along the cubic <001> axes. As both, shear and strain, are movements in different directions, it is not clear how the atoms move towards a modulated martensite. These aspects suggest that neither of the two concepts can sufficiently describe the transformational and functional properties of modulated martensite completely.

To address this deficiency and explain key functional properties of these Heusler alloys, we used large-scale density functional theory (DFT) calculations on the $Ni_2MnGa$ prototype system. We first briefly summarize the geometry and constraints of adaptive martensite. Then, we sought to verify the existence of an interaction between nanotwin boundaries and determined the oscillating interaction energy. We investigated the formation of the so-called *a/b* twin boundaries at the mesoscale to explore the effects of the interaction energy on the hierarchical martensitic microstructure. We also explored the role of nanotwin ordering in the hysteresis during a martensitic transition. Following this, we sought to understand the origin of the oscillating interaction energy and the stability of different modulations. Finally, we analysed the relaxation of the atoms within the lattice to understand the connection between soft phonons and nanotwinning.

## 1. Adaptive nanotwinning in a nutshell

In Fig. 1, we sketch how austenite (A) and different modulated martensites (M) are related through the central constraint of length conservation. Minimising the elastic energy at the phase boundary requires that the cubic lattice of austenite must be matched by the twinned martensite. For instance, the 14M martensite consists of simple, non-modulated (NM) tetragonal martensite twins with a $c/a|_{NM}$ = 1.26. Twin boundaries are introduced every two long ($n$) and five short ($\bar{m}$) lattice planes.[16] This arrangement of long and short axes retains an invariant, i.e. unstrained plane, the so-called habit plane that forms the phase boundary. It is favourable in terms of elastic energy; more details are described in Refs.[13-15,21] 10M martensite, discussed later in this paper, follows the same principle. The $c/a|_{NM}$ ratio is, however, reduced to 1.16 and twin boundaries are introduced after two and three lattice planes.[22] In case of a negligible twin boundary energy compared to the elastic energy, the modulation ($n+m$) should be the smallest possible number with *n* and *m* being integers.

Twinning can also occur within a unit cell; thus, it is appropriate to consider a martensitic "building block", which has just half the length and width of an ordered Heusler unit cell. In Zhdanov notation, these periodic arrangements are denoted as $(n\bar{m})_2$, where *n* and *m* are the number of tetragonal building blocks of different orientation, which are connected by $(101)_{NM}$ twin boundaries. If *n+m* is odd, the structure must be repeated to conform with chemical ordering. If *n* is unequal *m*, the unit cell becomes monoclinic.

The modulated structure is then represented as *2(n+m)*M. The stacking sequences corresponding to the most common modulations are $(2\bar{3})_2$ (10M) and $(2\bar{5})_2$ (14M). One observes larger *m* and thus larger $c/a|_{NM}$ at compositions with higher electron density and higher transformation temperatures.[23,24]

Conservation of length at the habit plane is approximately described by the equation $n \cdot c_{NM} + m \cdot a_{NM} = (n+m)b_M = (n+m)a_A$, where $b_M$ is commonly identified as the middle lattice constant of the modulated martensite of monoclinic crystal symmetry.[12] For a given tetragonal distortion $c/a|_{NM}$ and $a_A$, this equation can in general not be fulfilled by small integer numbers *n* and *m* but there is a small "remainder". Accordingly, the condition for an exact match at the habit plane $b_M = a_A$ is not fulfilled and there is a small misfit, which can be closed by faults in the *n-m* stacking sequence.[12] As a hierarchical twin-within-twins microstructure forms in martensite,[13,14,25] we will call the twin boundaries occurring at the atomic scale "nanotwin" boundaries in order to distinguish them from twin boundaries occurring at larger length scale.

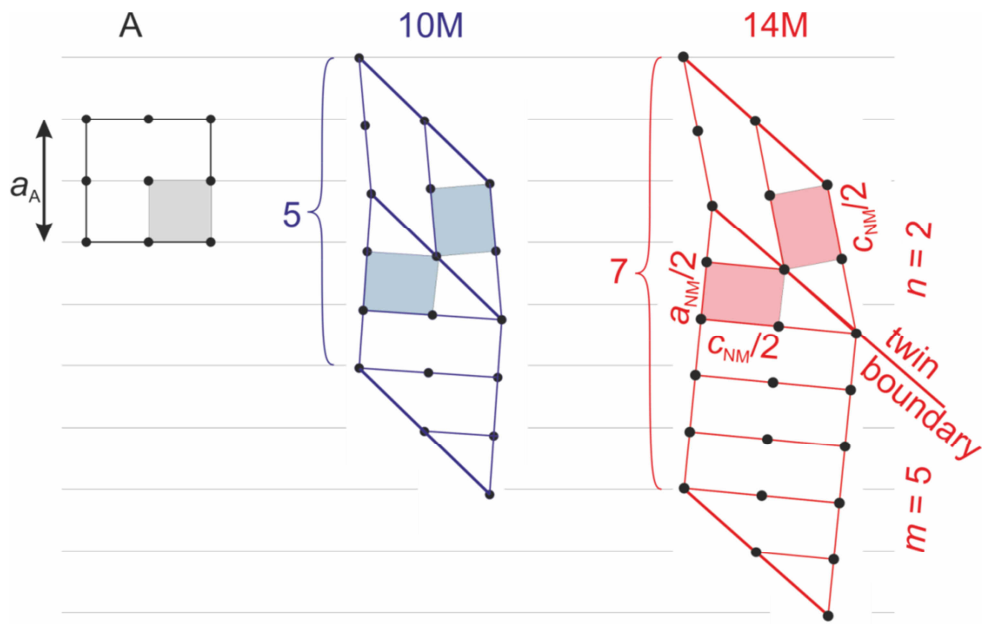

**Fig. 1: Geometry of austenite (A) and 10M and 14M modulations occurring in Ni-Mn-Ga with a $(2\bar{5})_2$ and $(2\bar{3})_2$ stacking, respectively.** The tetragonal NM martensitic and cubic austenitic building blocks are marked by coloured areas and their lattice parameters. To form an elastically compatible phase boundary with the austenite, the tetragonal martensite introduces twin boundaries at the nanoscale (thick lines), which mirror the building blocks, i.e. a quarter of the NM unit cell. For each modulation, a combination of *n* long and *m* short sides of the tetragonal building blocks keeps the lengths of austenite and modulated martensite invariant, as sketched by the horizontal lines. With increasing tetragonal distortion $c/a|_{NM}$ of the underlying tetragonal building block, a higher number *n+m* of tetragonal building blocks is required.

## Results and discussion

## 2. Oscillating interaction energy of nanotwins in dependency of their spacing

For many types of microstructural defects there is some kind of interaction, which results in the formation of new patterns. A well-known example from metal physics is the interaction between dislocations, resulting in the formation of small-angle grain boundaries. For twin boundaries, however, no such interaction is known yet. To prove the existence of an interaction between nanotwin boundaries, we used DFT to investigate the total twin boundary energy $\gamma_{tot}$. $\gamma_{tot}$ represents the excess energy for connecting two martensitic variants of a given width $n$ by a twin boundary divided by the interface area. Using the energy difference between a nanotwinned structure and the non-modulated reference (see Methods), we decomposed $\gamma_{tot}$ into two parts: 1) a bare formation energy of an isolated twin interface $\gamma_\infty$, connecting two variants of infinite length, and 2) the interaction energy $\gamma_n$ between neighbouring twin boundaries that are $n$ lattice planes apart.

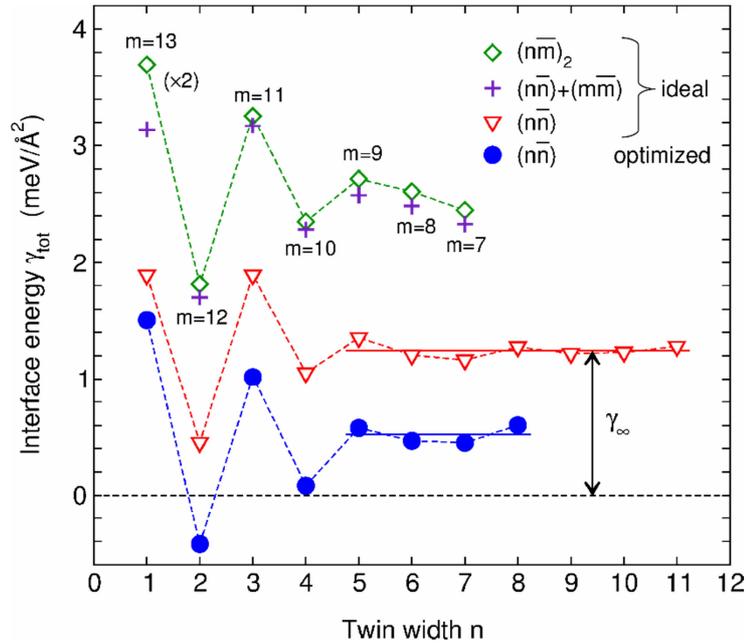

**Fig. 2: Oscillating interface energy of nanotwin boundaries $\gamma_{tot}$ separated by $n$ martensitic building blocks.** The $\gamma_{tot}$ of ideal symmetric $(n\bar{n})$ twins (triangles) was decomposed into the oscillations interaction of nanotwins, $\gamma_n$, at small $n$, and into a constant contribution $\gamma_\infty$ at large $n$. The latter describes the pure nanotwin boundary formation energy. The martensitic building blocks had a tetragonal distortion of $c/a|_{NM} = 1.26$. Asymmetric $(n\bar{m})_2$ arrangements (diamonds) could be predicted from the superposition of the respective $\gamma_{tot}$ obtained from the symmetric $(n\bar{n})$ and $(m\bar{m})$ cases (crosses), proving the additive relation. We used $n + m = 14$ building blocks where the twin boundaries were separated by $n$ and $m = 14 - n$ building blocks, respectively. As this geometry contains twice as many nanotwin boundaries compared to the symmetric one, the offset was 2 x $\gamma_\infty$. Upon structural optimisation (circles), $\gamma_{tot}$ decreased rather uniformly, i.e. independent of $n$, indicating that relaxations were confined to the interface.

All twins consisted of stoichiometric Ni$_2$MnGa with a tetragonal distortion of $c/a|_{NM}$ = 1.26, which is the ground state according to previous DFT calculations.[26-29] (see Supplementary Information for other $c/a|_{NM}$). In a first series, we considered a symmetric arrangement of twin boundaries $(n\bar{n})$ and varied the spacing of both twin boundaries from $n$ = 1 to 12 (Fig. 2). The triangles represent the case of ideal twin boundaries, where the atoms in each twin were fixed to positions obtained from a simple mirroring of the NM martensite. The corresponding $\gamma_{tot}$ showed a pronounced oscillatory pattern, which proved a short range interaction between nanotwin boundaries. For $n \geq 5$, oscillations diminished and thus $\gamma_\infty$ was obtained by averaging the total interface energies for $n \geq 5$. The decomposition of $\gamma_{tot}$ into interface and interaction contribution also holds if neighbouring twins are of arbitrary width $n$ and $m$. For $n > 1$, the corresponding $\gamma_n$ and $\gamma_m$ can be summed up (crosses in Fig. 2) to obtain the effective interaction energy of a $(n\bar{m})_2$ arrangement (diamonds). We thus suggest that it is sufficient to consider only the pairwise interaction energy and to neglect further terms.

Finally, we allowed for relaxation with respect to all four monoclinic parameters of the simulation box and atomic positions (circles). This resulted in a nearly constant shift by -0.8 meV/Å$^2$ for all $n$, thus $\gamma_\infty$ = 0.5 meV/Å$^2$ now included all structural relaxations caused by the placement of the twin boundary. The oscillating pattern did not change, proving that the interaction between nanotwins was stable with respect to the slight displacement of the atoms.

A pronounced minimum of the interaction energy was obtained for $n$ = 2. We found a substantial increase in interaction energy from $n$ = 2 to $n$ = 1 in agreement with the absence of modulation with $n$ = 1 in experiments. Compared to a disordered arrangement of nanotwin boundaries, which contains a broad distribution of $n$, arrangements with a maximum number of twins with $n$ = 2 were favourable in terms of the total energy. This explains why all experimentally observed ordered modulations contain double-layered twins.

Our disentangling of interaction and twin boundary energy gives a better understanding of the observed modulations compared to previous work, which considered only a positive twin boundary energy.[11,15] Surprisingly, after optimisation, we obtained a slightly negative $\gamma_{tot}$ for double layered twins, which become the global energy minimum. The optimum $(2\bar{2})$ stacking exhibits orthorhombic symmetry and is called 4O. Indeed, there are several experimental reports on the 4O phase in metamagnetic Heusler alloys, where 4O is always observed together with other modulations.[30,31] In a recent independent work, the geometry and electronic structure of the 4O phase is described in more detail.[32] We will discuss the origin of the negative $\gamma_{tot}$ for $n$ = 2 in section 6. Interface- and interaction energies between the twin boundaries determined here were within the range of thermal energies. Hence, thermal fluctuations will likely affect the arrangement of the twin boundaries when transforming to the modulated state of martensite.

### 3. Formation of $a_M/b_M$ twin boundaries by ordering nanotwins

What are the consequences of the interaction energy on the martensitic microstructure beyond the length scale of nanotwinning? Does the integer number of building blocks within a periodic modulation induce a new type of twin boundary at mesoscopic length scale? To find out, we reconsidered the slightly different length of a complete $(2\bar{5})_2$ modulation compared to seven austenitic unit cells along the habit plane. This misfit can be closed by stacking faults[12] that are inserted with a substantially larger period than the modulation period. As this stacking fault period is not solely defined by the lattice constants of martensite but also depends on the austenite, the arrangement is incommensurate when measuring a completely martensitic sample, which is in agreement with several diffraction experiments.[5,22,59] Compared to a periodic $(2\bar{5})_2$ stacking sequence, the presence of stacking faults increases the energy of an incommensurate nanotwin arrangement. In addition, one has to consider that the martensitic transformation of the technologically most interesting alloys occurs in vicinity or even well above room temperature. At finite temperatures, entropy further disturbs the periodic arrangement owing to the large number of possible configurations. For simplicity, we considered a disordered arrangement of nanotwins as a starting point, which was only constrained by the total $n/m$-ratio of all short vs. all long axes, adapting to the austenite. We proposed that this disordered state is an intermediate state between the propagating phase boundary connecting austenite and ordered martensite. It may also be an intermediate state with respect to time and temperature. In Fig. 3 we illustrate consequences of the transition from a disordered (a) to an ordered (b) arrangement of nanotwins. This transition was driven by minimization of the total interaction energy at low temperatures. Ordering proceeded by the movement of twin boundaries, which is easily possible by moving line defects that are known as disconnections across a twin boundary plane.[33] We identified two possible types of order – $(2\bar{5})_2$ and $(5\bar{2})_2$ – and were able to reveal the mirror symmetry between both kinds by highlighting the common modulated unit cells (in red in Fig. 3b). When using the lattice axis of the 14M unit cell,[21] the mirror plane exchanged $a_{14M}$ with $b_{14M}$. We infer that ordering of nanotwins result in the formation of $a_{14M}/b_{14M}$ twin boundaries between the different orientations of ordered modulated martensite.

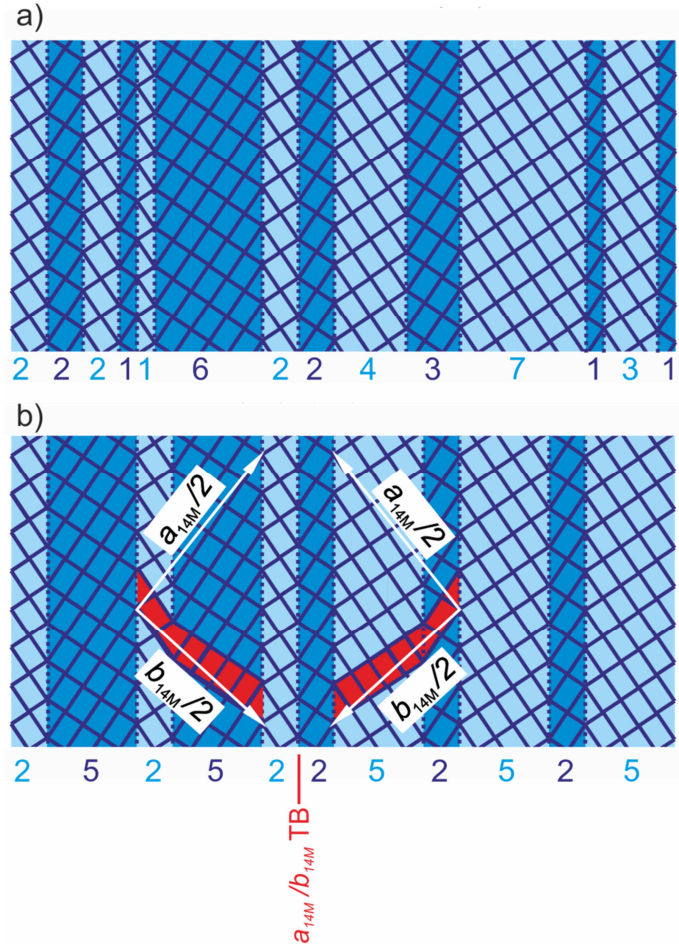

**Fig. 3: Schematic view how minimizing the total interaction energy results in a transition from a disordered arrangement of nanotwins to periodic 14M arrangements connected by mesoscopic $a_{14M}/b_{14M}$ twin boundaries.** a) Starting point was an irregular arrangement of nanotwin boundaries (dotted lines) connecting two different orientations of the tetragonal building blocks (marked in light and dark blue). b) Without changing the number of building blocks for each direction, an almost regular arrangement was possible, which minimised the interaction energy. This is illustrated for the 14M martensite exhibiting an ideal $(2\bar{5})_2$ stacking sequence. Though at both sides of the $a_{14M}/b_{14M}$ twin boundary (TB) a $(2\bar{5})_2$ stacking occurs, on the left side these are two light and five dark variants and on the right side five light and two dark ones. Thus, the orientation of the 14M unit cells (marked in red) is mirrored by the $a_{14M}/b_{14M}$ twin boundary.

In addition to their geometry, we could also predict the spacing of $a_{14M}/b_{14M}$ twin boundaries. Because a martensitic transition is diffusionless, ordering should not change the total number of unit cells. Moreover, ordering and the formation of $a_{14M}/b_{14M}$ twin boundaries should not alter the total number of each variant orientation because this would affect the length at the habit plane. On one hand, this fixed the length ratio of the variants $a_{14M}$ vs. $b_{14M}$. On the other hand, because $b_{14M} < a_A < a_{14M}$, the formation of $a_{14M}/b_{14M}$ twin boundaries allowed for a fine compensation of the remaining strain at the phase boundary, which is not compensated for by a fully ordered 14M structure. The difference between $a_{14M}$ and $b_{14M}$ lattice parameters[15] is 7% and thus introducing $a_M/b_M$ twin boundaries could compensate a substantial strain in order to minimise elastic energy. An $a_{14M}/b_{14M}$ twin boundary represented a stacking fault, requiring additional interaction energy compared to the ideal stacking sequence. As no further disturbance of the lattice occurs, the excess energy of an $a_M/b_M$ twin boundary is expected to be low and in the same order of the interaction energy. Because $a_{14M}/b_{14M}$ twin boundaries compensated substantial strain at very low twin boundary energy they could be introduced at high density, following the concept sketched in section 1. This agrees with recent reports of rather narrow spacings of these mesoscopic twin boundaries on 10M.[4,25]

The concept of ordering of nanotwins could explain some of the features of a martensitic microstructure up to the continuum length scale limit. In terms of the commonly observed hierarchical twin-within-twins microstructure,[11,13,14] $a_{14M}/b_{14M}$ twinning was found to represent the second smallest level of twinning. The $a_{14M}/b_{14M}$ twin boundaries were parallel to nanotwin boundaries, which is uncommon for two hierarchical levels of twin boundaries because this alignment does compensate the same strain component. The $a_{14M}/b_{14M}$ twin boundaries must thus have a different origin than common twinning. This is in agreement with the strain minimization and the interaction of nanotwin boundaries of the ordering process discussed above.

## 4. Ordering contributed to hysteresis losses

Does ordering nanotwins contribute to undesired macroscopic properties of a martensitic sample, such as hysteresis? Understanding and minimising hysteresis is crucial in order to use these materials in magnetocaloric refrigeration applications and as a conventional shape memory material. Hysteresis losses do not only reduce efficiency but may completely inhibit the reversibility in the low magnetic fields that is achievable by permanent magnets.

In a previous section, we established that ordering was driven by the minimization of the total interaction energy and this gain of energy was dissipated during cooling. The dissipated energy lost during a hysteresis amounted to 3.9 MJ/m³ during cooling for the interaction energies of 14M (Fig. 2, for details see Supplementary Information). This is a substantial fraction of the experimentally observed hysteresis loss of 16 MJ/m³, considering that also nucleation and volume change during a martensitic transformation contribute to the hysteresis. A similar amount of energy might be dissipated during the backward transformation, if the formation of an unstrained interface to the austenite also involves a disordering of the twin arrangement.

As a microscopic explanation for hysteresis, ordering nanotwins also provided the mechanism behind the so-called $\lambda_2 = 1$ approach, which predicts a minimum hysteresis when the middle eigenvalue of the transformation matrix approaches one.[34,35] At the continuum level, the approach describes the compatibility of a single austenite/martensite interface, which is required during a first order phase transformation because both phases can coexist. Each interface is a defect within the microstructure and thus requires an excess energy, which is dissipated after the complete transformation of the sample. The same dissipation occurs again during the reverse transformation, which relates to hysteresis. For adaptive martensites, the $\lambda_2 = 1$ criterion is fulfilled if $na_{NM} + mc_{NM} = (n+m)a_A$. If this criterion was met precisely, the material could transform directly to a $(2\bar{5})_2$ stacking and neither ordering nor $a_{14M}/b_{14M}$ twin boundaries were required. Deviations from this ideal situation increased the misfit, which increased disorder and thus hysteresis losses in agreement with the $\lambda_2 = 1$ approach.

However, the $\lambda_2 = 1$ approach must be expanded. It considers only the phase boundary, but not the volume transformed. As the two-dimensional phase boundary has a negligible contribution to the total energy compared to the volume, it does not allow quantifying hysteresis losses. In contrast, the volume was considered in the concept of ordering because the interaction energy was dissipated throughout the sample when ordering follows the phase boundary. Ordering nanotwins thus helps overcome the shortcoming of the $\lambda_2 = 1$ approach, and we regard it essential to understand hysteresis in modulated martensite.

### 5. Magnetic origin of the interaction energy

What is the microscopic origin of the oscillating interaction energy? In a nanotwinned structure we can expect that the charge density at twin interfaces is perturbed, which causes atoms to displace. In analogy to Friedel oscillations originating from a point defect, twin interfaces will result in a superposition of charge density waves, which can fix preferred distances for the next defect. However, the strong Coulomb interaction between electrons and ions should then lead to a considerable oscillatory displacements perpendicular to the interface, which we did not observe here (see section 7). In turn, charge-neutral oscillations can be expected from magnetic perturbations. Indeed, an oscillatory variation of the magnetic moments in the 6M, 10M and 14M along [101] has been reported in earlier DFT calculations.[36,37] Inspecting the magnetic moments in the nanotwinned structures (see Supplementary Information), we observed a significant decrease of about 7% in the magnetic moments of Ni atoms at the twin interface. In the adjacent lattice planes, a reduction of around 5% remained. The magnetisation has essentially recovered in the next-nearest plane, whereas only minor oscillations were visible further away. The effective interface width was thus below three lattice planes; interfaces overlapped only at a distance of *n* = 4 and below.

The Ni moments are stabilized by their exchange with the neighbouring Mn electrons. In turn, the magnetic exchange coupling between Mn atoms depends on the interatomic spacing.[38,39] Here, Ruderman-Kittel-Kasuya-Yosida-like oscillations are identified.[40-42] These result in alternating positive (ferromagnetic) and negative (antiferromagnetic) exchange interactions $J_{ij}$ between two atoms $i$ and $j$ of NM martensite, which we observed here (Fig. 4, left panel) in agreement with previous reports.[41,43] Significant antiferromagnetic interactions between the Mn atoms appear up to interatomic distances of 10.2 Å, which corresponds to approximately five (110) lattice planes with $d = 2.10$ Å. The right panel shows the interactions present in 4O, which exhibits a dense $(2\bar{2})$ stacking and was most favoured by interaction energy. Comparing the exchange constants of both structures, we could identify two competing trends. First, the ferromagnetic $J_{ij}$ of neighbouring Ni-Mn and Mn-Mn pairs of 4O decreased by about 1 meV when compared to NM, which destabilised the 4O. Conversely, we also found a much smaller number of frustrated (negative) antiferromagnetic interactions in the 4O compared to the NM. Summing up all Mn-Mn and Mn-Ni exchange constants over a distance from 2.5 to 19 Å finally yielded a substantial contribution of about 6 meV/f.u. in favour of the 4O, which was of similar magnitude as the total energy difference. This suggests that the twin interaction originates from frustrated antiferromagnetic exchange in the NM and explains the frequent observation of $n = 2$ in experiment.

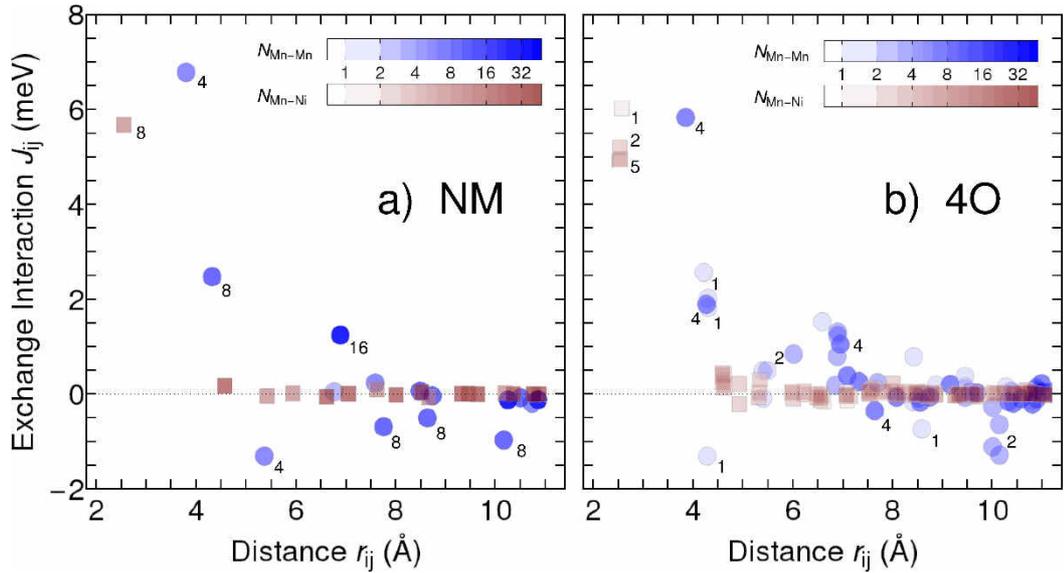

**Fig. 4: The magnetic origin of the interaction energy.** Magnetic exchange constants $J_{ij}$ within a) the tetragonal NM martensite with $c/a|_{NM}=1.25$ and b) the optimised 4O demonstrating the decrease of magnetic frustration in 4O. Exchange constants between pairs of Mn atoms are marked by blue circles and between Mn and Ni atoms by red squares. Positive values conform with the ferromagnetic order and negative values are frustrated and increase the energy. The saturation of the colours corresponds to the occurrence $N$ of the respective interaction relative to one formula unit $Ni_2MnGa$, indicating their impact on the total energy. For clarity, the multiplicity of selected interactions is marked explicitly by small numbers close to the symbols.

## 6. Twin boundary energy determined the phase sequence: A→10M→14M→NM/4O

Why do only particular types of modulations (as described in section 1) occur and not any other types? Are some modulations stabilised by a negative twin boundary energy rather than the common consideration of a twin boundary as a defect requiring excess energy? To understand which modulations are favoured, we calculated the energy difference of $(n\bar{m})_2$ nanotwinned arrangements with respect to an austenite unit cell in dependence of the tetragonal distortion of the underlying tetragonal building block $c/a|_{NM}$. Relaxation was allowed for all ions within the simulation cell, but change in cell size was not. We start the discussion of the results presented in Fig. 5 from the right-hand side of the figure. The $(2\bar{2})$ nanotwinning modulation, or 4O, was found to be lowest in energy, in accordance with Fig. 2. Its global energy minimum was located around $c/a|_{NM}$ = 1.25, which is about 5 meV/f.u. below the fully optimised NM martensite. This was still a rather small difference in energy, which depends on the magnetic state, as shown above. It was further diminished by 1 meV/f.u., when we took into account the zero-point energy of the lattice vibrations. Most important was that the 4O does not fulfil adaptivity[44], which means that 4O could not form directly, but only as a result of a complex, long-range ordering process. In other words, the kinetics (i.e. huge hysteresis) hindered the formation of 4O, an aspect that should be taken into account in addition to thermodynamics.

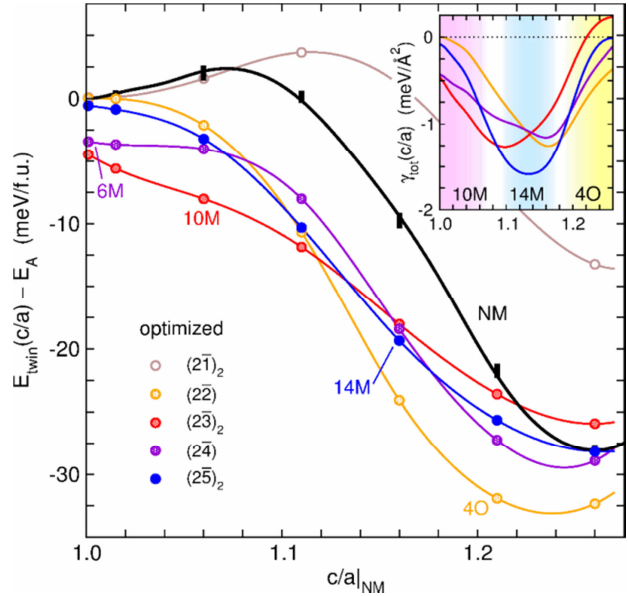

**Fig. 5: Nanotwinning provided an effective relaxation path, optimising the energy of martensite along the entire Bain path.** Comparison of the energy of NM martensite exhibiting tetragonal distortion $c/a|_{NM}$ (black line) with the corresponding relaxed $(n\bar{m})_2$ nanotwins (coloured circles). The values are given relative to the energy of austenite $E_A$. Because nanotwins were lower in energy than their NM building blocks, they formed an effective way for relaxation towards the martensite ground state. The inset shows the corresponding average twin boundary energy $\gamma_{tot}(c/a)$ calculated from the interpolation of the total energy in the main figure and the corresponding interface area and density of each $(n\bar{m})_2$ arrangement. It favoured the 10M for $c/a|_{NM} < 1.09$, the 14M for $1.09 < c/a|_{NM} < 1.18$ and the 4O at larger tetragonal distortions.

The $(2\bar{3})_2$, $(2\bar{4})_2$ and $(2\bar{5})_2$ modulations exhibited the next lowest energies. They were found to be almost degenerated, which agrees with the high density of stacking faults one observes in the 14M martensite.[45] At $c/a|_{NM}$ = 1.26 the energy of NM was equally low. Indeed, a microstructural transformation from nanotwinned martensite towards NM, which proceeds by the coarsening of twin boundaries has been reported.[13] This process requires a simultaneous annihilation of twin boundaries, which increases the energy barrier compared to the movement of a single nanotwin boundary required for the ordering. This is consistent with experiment, where a transformation from A to 14M and then to NM is often observed at large $c/a|_{NM}$ and low temperatures.[23,24]

For $c/a|_{NM}$<1.22, all modulations, except $(2\bar{1})_2$, were found to be significantly lower in energy than the intermediate NM building blocks from which they were constructed. This corresponded to a negative $\gamma_{TB}$ (Fig. 5, inset). This underlined - apart from the clear indication that nanotwins can be thermodynamically stable – that nanotwinning provides an effective relaxation channel towards an NM ground state. Our calculations revealed that the $(2\bar{5})_2$ modulation exhibits the lowest $\gamma_{tot}$. The 14M even outperformed the 4O, which had the lowest total energy, but a higher interface density. Because $\gamma_{tot}$ naturally varied with $c/a|_{NM}$, this eventually changed above $c/a|_{NM}$ = 1.18 in favour of the 4O, while for $c/a|_{NM}$ < 1.09, interfaces in a $(2\bar{3})_2$ sequence were preferred. The sequence of structures obtained was 10M→14M→NM or 4O with increasing $c/a|_{NM}$, which is in agreement with the compositional trend found in many Ni-Mn-based Heusler systems.[46]

For $c/a|_{NM}$→1, the energy did not approach zero for some structures. This is expected from the adaptive model where all twin arrangements should become equivalent. However, the energetic order of the modulations agrees well with other recent ab initio surveys, which consider soft phonons.[29,47] This indicates an independent source of lattice relaxation, i.e. the instability of the transverse acoustic (TA$_2$) phonon branch around the reciprocal lattice vector $q$ = 1/3 [101]. It is associated with the formation of the 6M premartensite[48-53] and originates from parallel nesting sheets of the Fermi surface.[9,10,54-56] The soft phonon results in a sinusoidal modulation of the atomic positions, which repeats once within three building blocks and is thus commensurate with a $(3\bar{3})$ or $(2\bar{4})$ twin structure. Despite the higher energy compared to 14M and NM martensites, the 6M and also the 10M are stabilized at finite temperatures by vibrational entropy.[6,57] The $(2\bar{3})_2$ and to a lesser extent the $(2\bar{5})_2$ modulation also benefit from a soft phonon. It is described by reciprocal vectors of $q$ = 2/5 [101] and $q$ = 2/7 [101], respectively, because the modulations need to complete a second period to account for the chemical order according to the odd number of lattice planes. In contrast, $(2\bar{1})_2$ and $(2\bar{2})$ stacking (4O) cannot benefit from the soft phonon. This made 10M the energetically most favoured structure for a low tetragonal distortion. At a $c/a|_{NM}$ ratio of 1.16 adaptivity is also fulfilled,[15] maximising compatibility to austenite. Thus, both requirements, i.e. phase stability and an easy transformation path, explained the formation of 10M and its presence beyond $c/a|_{NM}$ = 1.09.

## 7. Soft phonons facilitated the transition to a nanotwinned martensite

The decrease in energy with increasing $c/a|_{NM}$ for 14M (Fig. 5) implies that the tetragonality is an appropriate parameter to describe the transformation path towards modulated martensite. To understand this path better, we examined the relaxation in more detail (Fig. 6). Relaxation became small for $c/a|_{NM}$ > 1.16 and there was hardly any visible difference at $c/a|_{NM}$ = 1.21. Therefore, it was appropriate to use the concept of twinning at the atomic scale, as the adaptive concept postulates. Yet, at low $c/a|_{NM}$ ratios, substantial movements of complete lattice planes occurred in a sinusoidal rather than zig-zag manner. This suggests a physical picture where soft phonons within the austenite start a movement as a linear instability that ultimately ends in a nanotwinned martensite at large atomic displacements.

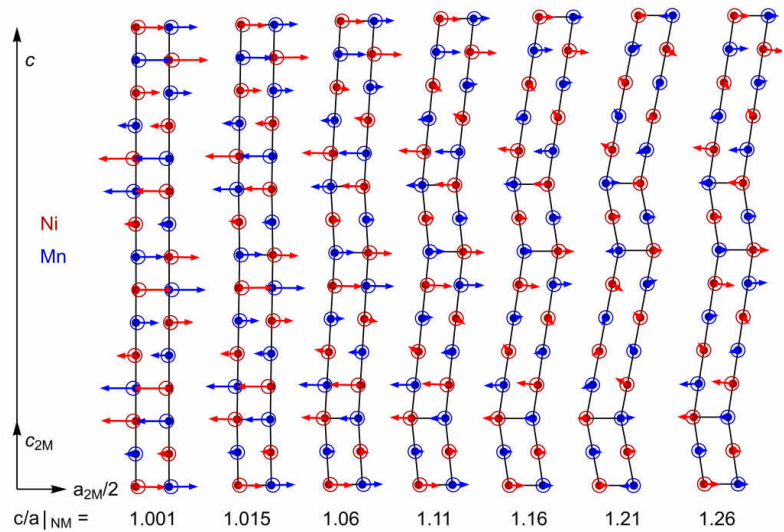

**Fig. 6): Transition of a soft phonon to nanotwinning with increasing tetragonal distortion $c/a|_{NM}$.** The positions of atoms within an ideal 14M modulation are depicted by solid circles while the relaxed positions are shown by open circles. Ni atoms are coloured in red and Mn atoms in blue. The Ga atoms are hidden beneath the Mn atoms and relaxed very similar. The arrows, illustrating the direction and the magnitude of the movement of the atoms, are enlarged by a factor of 5 for better visibility. We observed a sinusoidal pattern for small $c/a|_{NM}$ on the left-hand side according to the phonon instability, while only slight relaxations were found for the nanotwins on the right-hand side at the twin boundary. Relaxation perpendicular to the picture plane was negligible (less than 0.01% of a lattice constant). This projection uses the smallest unit cell of modulated martensite (2M) in contrast to the $L2_1$ description of Fig. 1.

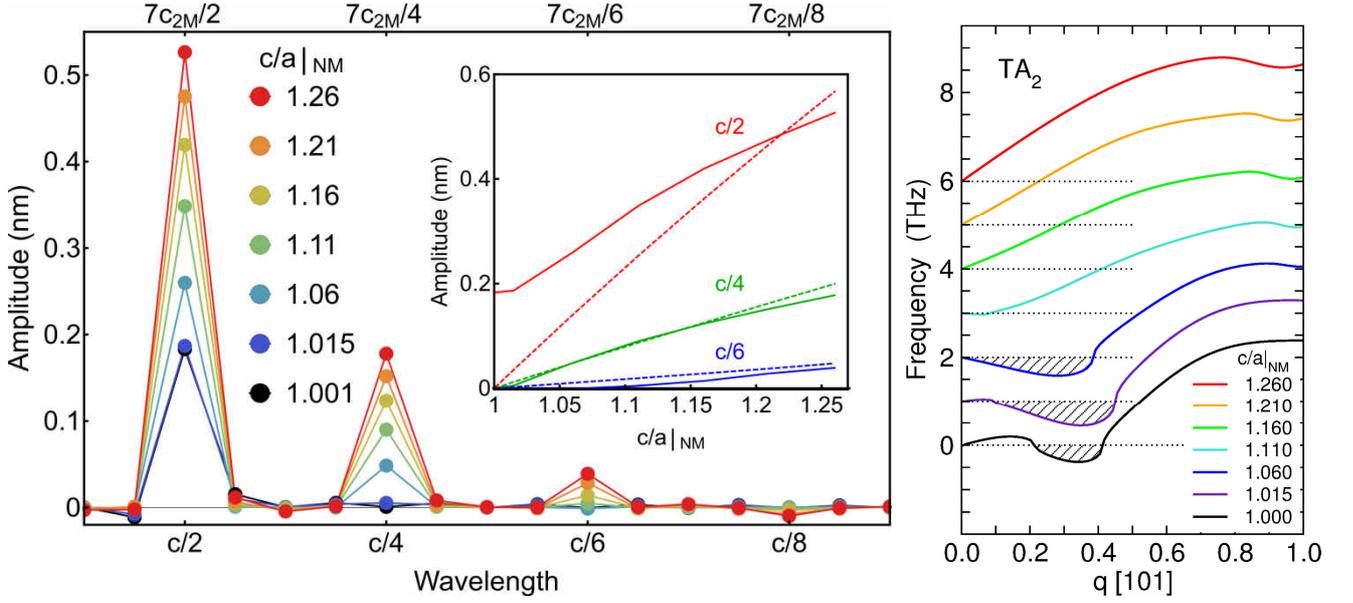

**Fig. 7. Quantifying the transition from a sinosoidal phonon instability to a zig-zag like nanotwinning.** Left: Discrete since Fourier transform of the relaxed lattice position in 14M. The coefficients describe the average amplitude of motion perpendicular to the modulation vector $c$. While the bottom axis gives the wavelength with respect to the full 2M-based unit cell of modulated martensite (long $c$-axis), the top one uses the reduced 2M description The coefficient $c/2$ corresponds to the first harmonic, describing a pure sine wave, while the higher order coefficients reproduced the zig-zag arrangement of the nanotwin boundaries. The inset shows the variation of the relaxed (solid lines) and unrelaxed coefficients (dashed lines) in dependence of $c/a|_{NM}$ demonstrating the gradual transition from a sinusoidal to a nanotwinned character. Right: TA$_2$ phonon branch in [101] direction for different $c/a|_{NM}$. For better visibility, the branches are consecutively shifted upwards by 1 meV and the imaginary parts of the dispersions (plotted as negative values) are presented as dashed areas. The data for $c/a|_{NM} = 1$ is taken from Ref.[58]

To characterize this transition from soft phonons to nanotwinning quantitatively, we calculated the discrete sine Fourier transformation coefficients of the relaxed atomic positions and compared them with the unrelaxed ones (Fig. 7, left; calculation details are described in the Supplementary Information). For low tetragonal distortions, the only dominant amplitude was found at a wavelength of $c/2$. With increasing $c/a|_{NM}$, the relative difference of the $c/2$ coefficients between the ideal and optimised twins was observed to decrease quickly. In contrast, the higher order coefficients were found to increase constantly, which was required to reproduce the zig-zag manner of nanotwinning. An identical behaviour was also observed for 10M (see Supplementary Information) and appears, therefore, to be of general nature.

The variation in $c/2$ at low $c/a|_{NM}$ can be understood from the TA$_2$ phonon branch in [110] direction (Fig. 7, right panel). As demonstrated previously,[58] first-principles calculations match the experimental phonon dispersion obtained from inelastic neutron scattering extremely well in this system. The imaginary frequencies of the TA$_2$ phonon in the cubic phase imply a dynamic instability favouring a sinusoidal modulation in the tetragonal building blocks. This anomaly is related to extend nesting of parallel sheets at the Fermi surface originating from non-bonding Ni-d$_{x^2-y^2}$ and d$_{3z^2-r^2}$ orbitals, which form a pronounced peak in the minority spin density of states right below the Fermi level. We observed that the region of imaginary frequencies in [101] direction widened slightly for very small tetragonal distortions. Here, soft phonons were found to stabilise 6M, 10M, and 14M - but not 4O, which would correspond to $q = 0.5$. With increasing $c/a|_{NM}$, the imaginary part was found to be smaller while the minimum shifts towards smaller $q$, i.e. longer wavelengths. Around $c/a|_{NM} = 1.1$, the non-bonding Ni-d$_{x^2-y^2}$ states pass the Fermi level, becoming unoccupied further on.[56] As a result, the dynamic instability vanished completely above $c/a|_{NM} = 1.11$, which suggests that the twinned structures were not supported by soft phonons for larger $c/a|_{NM}$. Instead, they became intrinsically stable, maintained by their favourable twin boundary energies.

Dynamic instability and nanotwinning decreased the energy of martensite, and our results showed that for small $c/a|_{NM}$ both effects can be present at the same time – they are not mutually exclusive. Our computational approach considered only phonons, which were commensurate with the underlying twin structure. This, however, was only a technical limitation. In principle, twinned structures can combine with incommensurate modulations observed at comparatively small tetragonal distortions.[59-61]

The transformation mechanism described here overcomes several shortcomings of the original Bain concept. The Bain path only considers elongation and compression along the cubic <001> axes whereas the soft phonon results in {101} shear instabilities. The observed relaxation mechanism illustrated how shearing can result in a tetragonal distortion when nanotwinning is considered. Moreover, shearing occurred only in one single set of planes and no doubleshearing, as postulated by Bogers and Burgers[62], was required. Therefore, we conclude that the concepts of adaptivity and ordering could resolve the major deficiency of the Bain concept already at the nanoscale, i.e. the absence of a habit plane.

## 8. Summary and conclusions

Among martensitic Heusler alloys only modulated structures exhibit outstanding magnetocaloric and magnetic shape memory properties. In this work, we solved the controversy about the origin of the modulations by unifying both alternative explanations: Phonon softening in the austenite state initialized the movement of lattice planes, which seamingless ended in a nanotwinned, adaptive martensite. Our analysis of the relaxation process illustrated that both, shear and tetragonal strain occur simultaneously during the transformation to a modulated martensite. We identified the interaction energy between nanotwin boundaries as driving force for ordered arrangements of nanotwins. Ordering explained why modulations are periodic and allowed predicting the sequence A→10M→14M→NM/4O of all experimentally observed modulations. In particular, the 10M martensite was stabilized by the negative twin boundary energy found at intermediate transformation steps. The 4O was favoured by its negative interaction energy, which originated from frustrated magnetic exchange interactions being removed by the twinning.

We demonstrated that ordering nanotwins has key implications on the functional properties of martensitic Heusler alloys. The energy dissipated during ordering was a significant contribution to macroscopic hysteresis losses. For magnetocaloric refrigeration, these losses must be as low as possible and thus we proposed to develop new alloy compositions with weak interaction energy. For magnetic shape memory alloys the formation of *a/b* twin boundaries by ordering explained a peculiar generation of twinning within the twin-within-twins hierarchy, which is decisive for the easy movement of mesoscopic twin boundaries. With these examples, ordering nanotwins demonstrated its versatility to explain key functional properties. As our work gave all relevant energy values, this will also enable a detailed understanding of the dynamics of the ordering process, which is decisive for premartensitic and strain glass phenomena.

## Computational methods

The twin interfaces were constructed systematically from non-modulated tetragonal martensite applying the phenomenological theory of martensite[63] as described in our previous work.[64,65] To comply with periodic boundary conditions for the DFT cell, an even number of complementary twin boundaries was always simulated. Spin-polarised density functional theory calculations were carried out using the VASP package[66,67] as described in Ref.[28] with a plane wave cut-off $E_{cut}$ = 460 eV, PBE exchange-correlation and projector augmented wave potentials with the basis $3d^{10}4s^{2}4p^{1}$ for Ga, $3p^{6}3d^{6}4s^{1}$ for Mn and $3p^{6}3d^{9}4s^{1}$ for Ni. Optimisation of atomic positions and lattice parameters were conducted until the energy difference between two consecutive steps was below 0.1 µeV. In this case, Brillouin zone integration was carried out with the Methfessel-Paxton broadening scheme with σ = 0.1 eV. Total energies were calculated using the tetrahedron method. Comparisons were made between supercells of the same number of atoms and similar extensions. The final k-mesh was chosen 16x12 in the twin plane and a division of 2 to 16 perpendicular to it, depending on the extension of the supercell in this direction. Phonon calculations were carried out via the so-called force constant approach using the VASP code in combination with the PHON utility[68] as described earlier.[57,58] For the NM martensite, we calculated the forces from displacements of single atoms in 4x4x4 supercell (2x2x2 for the 4O) of the distorted Ni$_2$MnGa primitive cell in opposite directions in combination with a 4x4x4 Monkhorst-Pack *k*-mesh (5x4x3 for the 4O). The magnetic exchange constants were calculated with the SPR-KKR multiple scattering code of Hubert Ebert.[69,70] For the NM, we used 2176 k-points in the irreducible Brillouin zone (192 for the 4O) and considered angular momenta up to $l_{max} = 4$ (*f*-states), using again the PBE exchange-correlation functional.

**Acknowledgements**

This work was supported by the Deutsche Forschungsgemeinschaft (DFG) via the SPP1239 www.MagneticShape.de and SPP1599 www.FerroicCooling.de and grant FA 453/8. The authors gratefully acknowledged the use of the Opterox, Cray XT6/m and MagnitUDE (DFG grant INST 20867/209-1) supercomputer systems of the Faculty of Physics, the Center of Computational Sciences and Simulation (CCSS) and the Zentrum für Informations- und Mediendienste (ZIM) of the University of Duisburg-Essen. M.E.G. would like to thank G. Schierning and R. Theissmann for fruitful discussions.

**Author contributions**

S.F. suggested that there should be an interaction energy, which started this work. M.E.G carried out the DFT calculations., R.N. analysed the relaxation mechanism and calculated the hysteresis losses under supervision of K.N., U.K.R. described the implications of the order-disorder transition. S.F., M.E.G, R.N., U.K.R., P.E., R.P. and K.N. contributed to the manuscript.

**Additional information**

Supplementary information is linked to the online version of the paper. Correspondence should be addressed to M.E.G or S.F.

**Competing financial interests**

The authors declare no competing financial interests.

# References

[1] Krenke, T. *et al.* Inverse magnetocaloric effect in ferromagnetic Ni–Mn–Sn alloys. *Nature Mat.* **4**, 450 (2005).

[2] Ullakko, K., Huang, J. K., Kantner, C., O'Handley, R. C. & Kokorin, V. V. Large magnetic-field-induced strains in $Ni_2MnGa$ single crystals. *Appl. Phys. Lett.* **69**, 1966 (1996).

[3] Srivastava, V., Chen, X. & James, R. D. Hysteresis and unusual magnetic properties in the singular Heusler alloy $Ni_{45}Co_5Mn_{40}Sn_{10}$. *Appl. Phys. Lett.* **97**, 014101 (2010).

[4] Straka, L. *et al.* Highly mobile twinned interface in 10M modulated Ni–Mn–Ga martensite: Analysis beyond the tetragonal approximation of lattice. *Acta Mat.* **59**, 7450 (2011).

[5] Schubert, M. *et al.* Collective Modes and Structural Modulation in Ni-Mn-Ga(Co) Martensite Thin Films Probed by Femtosecond Spectroscopy and Scanning Tunneling Microscopy. *Phys. Rev. Lett.* **115**, 076402 (2015).

[6] Dutta, B. *et al.* Ab initio Prediction of Martensitic and Intermartensitic Phase Boundaries in Ni-Mn-Ga. *Phys. Rev. Lett.* **116**, 025503 (2016).

[7] Devi, P., Singh, S., Manna, K. Suard, E.,Petricek, V. Felser, C., Pandey, D. Adaptive modulation in $Ni_2Mn_{1.4}In_{0.6}$ magnetic shape memory Heusler alloy, https://arxiv.org/abs/1611.06688.

[8] Heczko, O., Cejpek, P., Drahokoupil, J. & Holý, V. Structure and microstructure of Ni-Mn-Ga single crystal exhibiting magnetic shape memory effect analysed by high resolution X-ray diffraction. *Acta Mat.* **115**, 250 (2016).

[9] Lee, Y., Rhee, J. Y. & Harmon, B. N. Generalized susceptibility of the magnetic shape-memory alloy $Ni_2MnGa$. *Phys. Rev. B* **66**, 054424 (2002).

[10] Bungaro, C., Rabe, K. M. & Dal Corso, A. First-principles study of lattice instabilities in ferromagnetic $Ni_2MnGa$. *Phys. Rev. B* **68**, 134104 (2003).

[11] Niemann, R. *et al.* The Role of Adaptive Martensite in Magnetic Shape Memory Alloys. *Adv. Eng. Mat.* **14**, 562 (2012).

[12] Khachaturyan, A. G., Shapiro, S. M. & Semenovskaya, S. Adaptive phase formation in martensitic transformation. *Phys. Rev. B* **43**, 10832 (1991).

[13] Kaufmann, S. *et al.* Modulated martensite: why it forms and why it deforms easily. *New J. of Phys.* **13**, 053029 (2011).

[14] Müllner, P. & King, A. H. Deformation of hierarchically twinned martensite. *Acta Mat.* **58**, 5242 (2010).

[15] Kaufmann, S. *et al.* Adaptive Modulations of Martensites. *Phys. Rev. Lett.* **104**, 145702 (2010).

[16] Pons, J., Chernenko, V. A., Santamarta, R. & Cesari, E. Crystal structure of martensitic phases in Ni–Mn–Ga shape memory alloys. *Acta Mat.* **48**, 3027 (2000).

[17] Diestel, A., Backen, A., Rößler, U. K., Schultz, L. & Fähler, S. Magnetic domain pattern in hierarchically twinned epitaxial Ni–Mn–Ga films. *Appl. Phys. Lett.* **99**, 092512 (2011).

[18] Heczko, O., Kopecký, V., Sozinov, A. & Straka, L. Magnetic shape memory effect at 1.7 K. *Appl. Phys. Lett.* **103**, 072405 (2013).

[19] Chmielus, M., Chernenko, V. A., Knowlton, W. B., Kostorz, G. & Müllner, P. Training, constraints, and high-cycle magneto-mechanical properties of Ni-Mn-Ga magnetic shape-memory alloys. *Eur. Phys. J.-Spec. Top.* **158**, 79 (2008).

[20] Bain, E.C. The nature of martensite. *Trans. Am. Insitute Min. Metall. Eng.* **70**, 25–46 (1924).

[21] Niemann, R. & Fähler, S. Geometry of Adaptive Martensite, J. Alloys. Comp. submitted, http://arxiv.org/abs/1611.02535 (2016).

[22] Righi, L., Albertini. F., Pareti, L., Paoluzi, A. & Calestani, G. Commensurate and incommensurate "5M" modulated crystal structures in Ni–Mn–Ga martensitic phases. *Acta Mat.* **55**, 5237 (2007).

[23] Heczko, O., Scheerbaum, N. & Gutfleisch, O. in *Nanoscale Magnetic Materials and Applications* (eds. Liu, J. P. et al.) Ch. 14 (Springer, Dordrecht, 2009).

[24] Çakır, A. *et al.* Extended investigation of intermartensitic transitions in Ni-Mn-Ga magnetic shape memory alloys: A detailed phase diagram determination. *J. Appl. Phys.* **114**, 183912 (2013).

[25] Seiner, H., Straka, L. & Heczko, O. A microstructural model of motion of macro-twin interfaces in Ni–Mn–Ga 10M martensite. *J. Mech. Phys. Sol.* **64** 198 (2013).

[26] Ayuela, A., Enkovaara, J., Ullakko, K. & Nieminen, R. Structural properties of magnetic Heusler alloys. *J. Phys. Condens. Matter* **11**, 2017 (1999).

[27] Godlevsky, V. & Rabe, K. Soft tetragonal distortions in ferromagnetic $Ni_2MnGa$ and related materials from first principles. *Phys. Rev. B* **63**, 134407 (2001).

[28] Entel, P. *et al.* Modelling the phase diagram of magnetic shape memory heusler alloys. *J. Phys. D: Appl. Phys.* **39**, 865 (2006).

[29] Hickel, T. *et al.* Ab initio-based prediction of phase diagrams: Application to magnetic shape memory alloys. *Adv. Eng. Mater.* **14**, 547 (2012).

[30] Sutou, Y. *et al.* Magnetic and martensitic transformations of NiMnX(X=In,Sn,Sb) ferromagnetic shape memory alloys. *Appl. Phys. Lett.* **85**, 4358 (2004).

[31] Brown, P. J. *et al.* The magnetic and structural properties of the magnetic shape memory compound $Ni_2Mn_{1.44}Sn_{0.56}$. *J. Phys. Condens. Matter* **18**, 2249 (2006).

[32] Zelený, M., Straka, L., Sozinov, A. & Heczko, O. Ab initio prediction of stable nanontwin doublelayers and 4O structure in $Ni_2MnGa$. *Phys. Rev. B* **94**, 224108 (2016).

[33] Pond, R. C., Ma, X., Hirth, J. P. & Mitchell, T. E. Disconnections in simple and complex structures. *Phil. Mag.* **87**, 5289 (2007).

[34] Cui, J. *et al.* Combinatorial search of thermoelastic shape-memory alloys with extremely small hysteresis width. *Nat. Mater.* **5**, 286 (2006).

[35] Zarnetta, R. *et al.* Identification of Quaternary Shape Memory Alloys with Near-Zero Thermal Hysteresis and Unprecedented Functional Stability. *Adv. Funct. Mat*. **20**, 1917 (2010).

[36] Xu, N. *et al.* Oscillation of the magnetic moment in modulated martensites in $Ni_2MnGa$ studied by ab initio calculations. *Appl. Phys. Lett.* **100**, 084106 (2012).

[37] Bai, J. *et al.* First-principles investigation of magnetic property and defect formation energy in Ni-Mn-Ga ferromagnetic shape memory alloy. *Int. J. Quant. Chem.* **113**, 847 (2013).

[38] Khovaylo, V. V. *et al.* Magnetic properties of $Ni_{50}Mn_{34.8}In_{15.2}$ probed by Mössbauer spectroscopy. *Phys. Rev. B* **80**, 144409 (2009).

[39] Kainuma, R. & Umetsu, R. Y. in *Disorder and Strain-Induced Complexity in Functional Materials* (eds. Kakeshita, T. *et al.*) Ch. 3, (Springer, Dordrecht, 2012).

[40] Sasioglu, E., Sandratskii, L. & Bruno, P. Role of conduction electrons in mediating exchange interactions in Mn-based Heusler alloys. *Phys. Rev. B* **77**, 064417 (2008).

[41] Galanakis, I. & Sasioglu, E. Variation of the magnetic properties of $Ni_2MnGa$ Heusler alloy upon tetragonalization: A first-principles study. *J. Phys. D: Appl. Phys.* **44**, 235001 (2011).

[42] Entel, P., Gruner, M. E., Comtesse, D. & Wuttig, M. Interaction of phase transformation and magnetic properties of Heusler alloys: A density functional theory study. *JOM* **65**, 1540 (2013).

[43] Buchelnikov, V. D. *et al.* First-principles and Monte Carlo study of magnetostructural transition and magnetocaloric properties of $Ni_{2+x}Mn_{1-x}Ga$. *Phys. Rev. B* **81**, 094411 (2010).

[44] The adaptive equation reduces to $2a_{NM}+2c_{NM}=4a_A$. Together with volume conservation $a^2_{NM}c_{NM}=a^3_A$ there is only the trivial solution $c/a|_{NM}=1$.

[45] Y. Ge, N. Zarubova, O. Heczko and S-P. Hannula, Acta Mat. **90,** 151 (2015).

[46] Planes, A. Mañosa, L. & Acet, M. Magnetocaloric effect and its relation to shape-memory properties in ferromagnetic Heusler alloys. *J. Phys. Condens. Mat.* **21**, 233201 (2009).

[47] Gruner, M. E., Fähler, S. & Entel, P. Magnetoelastic coupling and the formation of adaptive martensite in magnetic shape memory alloys. *Phys. Stat. Sol. (b)* **251**, 2067 (2014).

[48] Zheludev, A. *et al.* Phonon anomaly, central peak, and microstructures in $Ni_2MnGa$. *Phys. Rev. B* **51**, 11310 (1995).

[49] Zheludev, A., Shapiro, S., Wochner, P. & Tanner, L. Precursor effects and premartensitic transformation in $Ni_2MnGa$. *Phys. Rev. B* **54**, 15045 (1996).

[50] Stuhr, U., Vorderwisch, P., Kokorin, V. & Lindgard, P. A. Premartensitic phenomena in the ferro- and paramagnetic phases of $Ni_2MnGa$. *Phys. Rev. B* **56**, 14360 (1997).

[51] Mañosa, L., *et al.* Phonon softening in Ni-Mn-Ga alloys. *Phys. Rev. B* **64**, 024305 (2001).

[52] Zayak, A. T., Entel, P., Enkovaara, J., Ayuela, A. & Nieminen R. M. First-principles investigation of phonon softenings and lattice instabilities in the shape-memory system. *Phys. Rev. B* **68**, 132402 (2003).

[53] Shapiro, S. M., Vorderwisch, P., Habicht, K., Hradil, K. & Schneider, H. Observations of phasons in magnetic shape memory alloy $Ni_2MnGa$. *Europhys. Lett.* **77**, 56004 (2006).

[54] Opeil, C. P. *et al.* Combined experimental and theoretical investigation of the premartensitic transition in $Ni_2MnGa$. *Phys. Rev. Lett.* **100**, 165703 (2008).

[55] Haynes, T. D. *et al.* Positron annihilation study of the fermi surface of $Ni_2MnGa$. *New J. Phys.* **14**, 035020 (2012).

[56] Siewert, M. *et al.* A first-principles investigation of the compositional dependent properties of magnetic shape memory Heusler alloys. *Adv. Eng. Mater.* **14**, 530 (2012).

[57] Uijttewaal, M. A., Hickel, T., Neugebauer, J., Gruner, M. E. & Entel, P. Understanding the phase transitions of the $Ni_2MnGa$ magnetic shape memory system from first principles. *Phys. Rev. Lett.* **102**, 035702 (2009).

[58] Ener, S. *et al.* Effect of temperature and compositional changes on the phonon properties of Ni-Mn-Ga shape memory alloys. *Phys. Rev. B* **86**, 144305 (2012).

[59] Righi, L. *et al.* Incommensurate modulated structure of the ferromagnetic shape-memory $Ni_2MnGa$ martensite. *J. Solid State Chem.* 179 3525 (2006).

[60] Singh, S., Bednarcik, J., Barman, S. R., Felser, C. & Pandey, D. Premartensite to martensite transition and its implications for the origin of modulation in $Ni_2MnGa$ ferromagnetic shape-memory alloy. *Phys. Rev. B* **92**, 054112 (2015).

[61] Mariager, S. O. *et al.* Structural and magnetic dynamics in the magnetic shape-memory alloy $Ni_2MnGa$. *Phys. Rev. B* **90**, 161103(R) (2014).

[62] Bogers, A.J. & Burgers, W.G. Partial dislocations on the {110} planes in the BCC lattice and the transition of the FCC into the BCC lattice. *Acta Metall.* 12, 255 (1964).

[63] Bhattacharya, K. *Microstucture of Martensite*, Oxford University Press, Oxford (2003).

[64] Gruner, M. E., Entel, P., Opahle, I. & Richter, M. Ab initio investigation of twin boundary motion in the magnetic shape memory Heusler alloy $Ni_2MnGa$. *J. Mater. Sci.* **43**, 3825 (2008).

[65] Gruner, M. E. & Entel, P. Simulating functional magnetic materials on supercomputers, *J. Phys. Condens. Matter* **21**, 293201 (2009).

[66] Kresse, G. & Furthmüller, J. Efficient iterative schemes for ab initio total-energy calculations using a plane-wave basis set. *Phys. Rev. B* **54**, 11169 (1996).

[67] Kresse, G. & Joubert, D. From ultrasoft pseudopotentials to the projector augmented-wave method. *Phys. Rev. B* **59**, 1758 (1999).

[68] Alfè, D. Phon: A program to calculate phonons using the small displacement method. *Comp. Phys. Commun.* **180**, 2622 (2009).

[69] Ebert, H. *et al*. The Munich SPR-KKR package. http://olymp.cup.uni-muenchen.de/ak/ebert/SPRKKR.

[70] Ebert, H. in *Electronic Structure and Physical Properties of Solids* (ed. Dreyssé), H., Springer, Berlin (2000).